\documentclass[twocolumn,english,prl]{revtex4}
\usepackage[]{fontenc}
\usepackage[latin1]{inputenc}
\usepackage{amsmath}

\makeatletter

\providecommand{\LyX}{L\kern-.1667em\lower.25em\hbox{Y}\kern-.125emX\@}

\usepackage{babel}
\makeatother
\begin{document}

\title{Reply to Comment on: {}``Radiation-Induced 'Zero-Resistance State'
and the Photon Assisted Transport''}

\author{Junren Shi$^{1}$ and X.C. Xie$^{2,\, 3}$}

\affiliation{$^{1}$Solid State Division, Oak Ridge National Laboratory, Oak Ridge,
Tennessee 37831\\
 $^{2}$Department of Physics, Oklahoma State University, Stillwater,
Oklahoma 74078\\
 $^{3}$International Center for Quantum Structures, Chinese Academy
of Sciences, Beijing 100080, China}

\begin{abstract}
We show that the comment by A.F. Volkov~\cite{Volkov2003} ignores
a delicate issue in the conductance measurement for a hall bar system.
In such system, $\rho _{xx}\approx \rho _{xy}^{2}\sigma _{xx}$ while
$\sigma _{xy}\gg \sigma _{xx}$, as correctly pointed out in Ref.~\onlinecite{Durst2003}.
We clarify that the so called {}``zero resistance state'' is actually
a {}``zero conductance state''. A discussion concerning the phase
transition induced by the negative conductance is presented.
\end{abstract}
\maketitle
In a comment~\cite{Volkov2003} to our paper titled {}``Radiation-Induced
'Zero-Resistance State' and the Photon Assisted Transport'',~\cite{Shi2003}
Volkov concludes that our work as well as the calculation in Ref.~\onlinecite{Durst2003}
can only produce the zero conductance instead of zero resistance.
This conclusion is certainly true for a normal transport measurement
without the presence of a magnetic field, where $\rho _{xx}=1/\sigma _{xx}$.
However, in the specific experimental setup we are talking about,
namely, a hall bar system,~\cite{Mani2002,Zudov2001,Zudov2002} the
relation between conductance and resistance is delicate albeit well
known:\begin{eqnarray*}
\rho _{xx} & = & \frac{\sigma _{xx}}{\sigma _{xx}^{2}+\sigma _{xy}^{2}}\, ,\\
\rho _{xy} & = & \frac{\sigma _{xy}}{\sigma _{xx}^{2}+\sigma _{xy}^{2}}\, .
\end{eqnarray*}
As correctly pointed out in Ref.~\onlinecite{Durst2003}, those experiments
lie in regime with $\sigma _{xy}\gg \sigma _{xx}$. This yields,\begin{eqnarray*}
\rho _{xx} & \approx  & \rho _{xy}^{2}\sigma _{xx}\, ,\\
\rho _{xy} & \approx  & 1/\sigma _{xy}\, .
\end{eqnarray*}
As the result, the measurement of longitudinal conductance is equivalent
to the measurement of longitudinal resistance. Our theory actually
suggest that the so called {}``zero resistance state'' is actually
a {}``zero conductance state'' . This has important implication
for a general non-hall bar system, \emph{i.e.}, we can only see radiation
induced {}``zero conductance'' instead of {}``zero resistance''.

When system enters into negative conductance (or negative resistance
for the S-shape CVC) regime, the instability will induce a phase transition
and the system will be stabilized in a new phase. It is very important
to study the detailed structure of such a new phase. On the other
hand, we stress that these are two separate issues: (a) the origin
of negative conductance (resistance); and (b) the structure of new
phase induced by the instability. These two issues are determined
by different sets of parameters. It becomes clear in our toy model:~\cite{Shi2003}
while the occurrence of negative conductance is fully determined by
the radiation power and electron density of states, the determination
of new voltage domain structure (see Fig.~4 in Ref.~\onlinecite{Shi2003})
requires more parameters such as barrier thickness, lead configurations,
dielectric constants, etc. It suggests that the structure of new phase
will be very sensitive to the special setup of systems. So it will
not be a surprise to find more exotic phases other than those discussed
in current stage of theoretical studies~\cite{Andreev2003,Anderson2003}
and mentioned in the comment.~\cite{Volkov2003}

Finally, we point out that while the transport anomaly is well known
in artificial systems,~\cite{Keay1995} the observation of similar
effect in a uniform system is still very interesting and striking.
Especially, the phase transition induced by the negative conductance
instability in such a uniform system is far from fully understood.
We conclude that the phase transition induced by the negative conductance
and the structure of resulting phase are the leading issues requiring
more theoretical and experimental pursuits.

\end{document}